# Exploring interfacial exchange coupling and sublattice effect in heavy metal/ferrimagnetic insulator heterostructures using Hall measurement, XMCD and neutron reflectometry

Qiming Shao[1*#], Alexander Grutter[2#], Yawen Liu[3], Guoqiang Yu[1,4], Chao-Yao Yang[1], Dustin A. Gilbert[2,5], Elke Arenholz[6], Padraic Shafer[6], Xiaoyu Che[1], Chi Tang[3], Mohammed Aldosary[3,9], Aryan Navabi[1], Qing Lin He[1,7], Brian J. Kirby[2], Jing Shi[3], Kang L. Wang[1,8*]

[1] Department of Electrical and Computer Engineering, University of California, Los Angeles, California 90095, USA

[2] NIST Center for Neutron Research, National Institute of Standards and Technology, Gaithersburg, Maryland 20899-6102, USA

[3] Department of Physics and Astronomy, University of California, Riverside, California 92521, USA

[4] Beijing National Laboratory for Condensed Matter Physics, Institute of Physics, Chinese Academy of Sciences, Beijing 100190, China

[5] Department of Materials Science and Engineering, University of Tennessee, Knoxville, Tennessee, 37996, USA

[6] Advanced Light Source, Lawrence Berkeley National Laboratory, Berkeley, California 94720, USA

[7] International Center for Quantum Materials, School of Physics, Peking University, Beijing, 100871, China

[8] Department of Physics and Astronomy, Department of Materials Science and Engineering, University of California, Los Angeles, California 90095, USA

[9] Department of Physics and Astronomy, King Saud University, Riyadh 11451, Saudi Arabia

[#]These authors contributed to this work equally.

*Corresponding authors: sqm@ucla.edu, wang@seas.ucla.edu

## ABSTRACT

We use temperature dependent Hall measurements to identify contributions of spin Hall, magnetic proximity, and sublattice effects to the anomalous Hall signal in heavy metal/ferrimagnetic insulator heterostructures with perpendicular magnetic anisotropy. This approach enables detection of both the magnetic proximity effect onset temperature and magnetization compensation temperature and provides essential information regarding the interfacial exchange coupling. Onset of a magnetic proximity effect yields a local extremum in the temperature dependent anomalous Hall signal, which occurs at higher temperature as magnetic insulator thickness increases. This magnetic proximity effect onset occurs at much higher temperature in Pt than W. The magnetization compensation point is identified by a sharp anomalous Hall sign change and divergent coercive field. We directly probe the magnetic proximity effect using X-ray magnetic circular dichroism and polarized neutron reflectometry, which reveal an antiferromagnetic coupling between W and magnetic insulator. At last, we summarize the exchange coupling configurations and the anomalous Hall effect sign of the magnetized heavy metal in various heavy metal/magnetic insulator heterostructures.

# I. INTRODUCTION

Like magnetic metals, ferrimagnetic insulators (FMIs) enable information storage and propagation through magnetization direction and spin wave transport, respectively. Unlike metallic systems, however, spin currents in FMIs do not require a commensurate charge transport component and thus are free of current-induced Joule heating, a beneficial feature for low power spintronic applications [1]. However, the electrical readout of magnetization and spin waves in FMIs have been challenging until the recent discovery of the inverse spin Hall effect (SHE) [2]. The inverse SHE in a heavy metal (HM) layer allows conversion from magnon spin current to charge current at the HM-FMI interface. In addition, the combined action of SHE and inverse SHE can give rise to a spin Hall magnetoresistance and anomalous Hall effect (AHE) [3, 4] (Fig. 1a). Interestingly, the sign of AHE in some HM/FMI systems can be tuned by varying the temperature [5-8]. Studies on the temperature dependence of magnetoresistance [9] and the AHE [7] have suggested the important role of the magnetic proximity effect (MPE), which appears below an onset temperature ($T_{\mathrm{on,MPE}}$) and induces a spontaneous magnetization in the interfacial HM layer. The magnetized HM produces an AHE (Fig. 1b), the sign of which may be different from that due to the SHE. Currently, a great deal of important information about the MPE, such as the onset temperature and whether ferromagnetic or antiferromagnetic exchange coupling is preferred, must be investigated by using spectroscopic or scattering techniques, such as X-ray magnetic circular dichroism (XMCD) and polarized neutron reflectometry (PNR), which require large facilities to implement.

Another important feature of FMIs is that they consist of multiple antiferromagnetically coupled magnetic sublattices, leading to a high characteristic frequency which is essential for high-speed spintronic applications [1]. In some cases, the different temperature dependencies of the sublattice magnetizations cause a magnetization compensation temperature ($T_{\mathrm{M}}$), at which the net magnetization is zero. The $T_{\mathrm{M}}$ is typically characterized using a bulk volume sensitive magnetometer, such as superconducting quantum interference devices. To probe local $T_{\mathrm{M}}$ in ultrathin FMI films, an alternative method is required. Although the AHE has been used as a local probe to detect $T_{\mathrm{M}}$ in ferrimagnetic metals [10, 11], it cannot directly probe an insulating system. As described above, by combining a HM with a FMI, the magnon spin current from the FMI, spin Hall magnetoresistance and AHE can be measured through inverse SHE. While the magnon spin current excited by the spin Seebeck effect [12] and spin Hall magnetoresistance [13] have been used to probe the $T_{\mathrm{M}}$, the AHE remains an unexplored avenue.

In this work, we demonstrate that the AHE provides an electrical desktop microprobe for detecting and separating AHE contributions, SHE, MPE, and sublattice orientation, in thin film bilayers consisting of tungsten (W) or platinum (Pt) and FMI thulium iron garnet ($Tm_3Fe_5O_{12}$, TmIG) or terbium iron garnet ($Tb_3Fe_5O_{12}$, TbIG). The observation of a local extremum in the AHE temperature dependence allows us to identify $T_{on,MPE}$, which increases with TmIG thickness and is much higher in Pt than W. The $T_M$ is identified by a sudden AHE sign change commensurate with a divergent coercive field ($B_C$). To confirm this interpretation, we directly probe the MPE using XMCD and PNR, which indicate antiferromagnetic exchange coupling between the W and the TmIG. Our data suggest that the Fe sublattice dominates the interfacial exchange coupling. These results provide a comprehensive picture of interfacial exchange coupling and sublattice effects in HM/FMI bilayers, which can be utilized in applications based on spintronics[14-17], magnonics[2], and spin caloritronics[18].

## II. MATERIALS

All TmIG(111) films were grown on $Nd_3Ga_5O_{12}$ (111) by pulsed laser deposition [19]. The TmIG films were grown at a moderate temperature of ~200°C by KrF excimer laser pulses of 248 nm in wavelength with a power of 150 mJ at a repetition of 1 Hz under 1.5-mtorr oxygen pressure with 12 wt. % ozone. Rapid thermal annealing processes were performed at 800°C for 5 min to magnetize the TmIG films. Each film has a nominal area 5 mm × 5 mm. We deposited W(5 nm)/MgO(2 nm)/$TaO_x$(3 nm) and Pt(5 nm) layers on top of TmIG using magnetron sputtering. For TmIG thicknesses 3 nm, 6 nm, 9 nm, 12 nm and 15 nm, W and Pt thin films each cover 2.5 mm × 5 mm. For other TmIG thicknesses, only W thin films are deposited on the TmIG. We also prepare the W/TbIG and Pt/TbIG thin films with detailed structures: GGG(111)/TbIG(6 nm)/W(5 nm)/MgO(2 nm)/ $TaO_x$(3 nm) and GGG(110)/TbIG(6 nm)/Pt(5 nm). The growth recipe for TmIG and TbIG thin films are the same.

## III. HALL MEASUREMENT

The HM/FMI thin films were patterned into Hall bar devices by using standard photolithography and dry etching for the four-probe lock-in resistance measurements. The magnetic field and temperature control were performed with a physical property measurement system.

### A. ONSET TEMPERATURE OF MAGNETIC PROXIMITY EFFECT

We first discuss contributions to the AHE and their temperature dependence, which allows detection of $T_{on,MPE}$. The MPE becomes pronounced when interfacial exchange coupling between the W and the TmIG is strong enough to suppress thermal fluctuations and induce a spontaneous magnetic moment in the interfacial HM layer. Magnetization induced by the MPE will give rise to an AHE, which we refer to as MPE-AHE (Fig. 1b). At higher temperature, thermal fluctuations dominate, disrupting the spontaneous W magnetization and eliminating the MPE-AHE. Even in the absence of the MPE, however, spin current transmitted across and reflected at the W/TmIG interface through the SHE and inverse SHE can give rise to an anomalous Hall signal [3], which we refer to as SH-AHE (Fig. 1a). A sign change or local extremum of the AHE may occur when a low-temperature MPE-AHE has the opposite sign of the SH-AHE which dominates at elevated temperatures.

To probe these contributions through transport measurements, we use $Nd_3Ga_5O_{12}$ (111)/TmIG ($t_{\mathrm{TmIG}}$)/(W, Pt)(5 nm)/MgO(2 nm)/$TaO_x$(3 nm), where $t_{\mathrm{TmIG}}$ is the TmIG thickness. We observe a clear AHE with a square hysteresis loop in the W/TmIG (Fig. 1c) due to the perpendicular magnetic anisotropy of TmIG thin films. In W/TmIG, the observed SH-AHE sign at room temperature is negative and the magnitude increases as temperature decreases from 360 K to 300 K due to increased spin mixing conductance [3, 20, 21]. As temperature is reduced further, we observe signatures of a MPE-AHE-related sign change in the W/TmIG (Fig. 1d). This behavior cannot be explained by a $T_{\mathrm{M}}$ since the $B_{\mathrm{C}}$ does not exhibit a divergent behavior (Fig. 1e). This suggests an emergent low-temperature MPE with an induced MPE-AHE with a positive sign. To understand how the MPE varies with temperature, we analyze the temperature dependence of the AHE resistance (Fig. 1e). Full AHE data in W/TmIG and Pt/TmIG are shown in Appendix A and B, respectively. As the temperature is reduced from above room temperature to low temperature (10 K), the anomalous Hall signal first increases in magnitude then decreases, with the extremum identified as $T_{\mathrm{ex}}$, before reversing sign. As the temperature is reduced, interfacial exchange dominates over the thermal fluctuations, stabilizing a MPE and contributing a positive AHE signal opposing the negative SH-AHE. Further, we note that MPEs are known to suppress the SHE and may reduce the spin mixing conductance [22]. Therefore, we expect an extremum near but somewhat below $T_{\mathrm{on,MPE}}$, which may then be used to indicate of $T_{\mathrm{on,MPE}}$ (Fig. 1e). Detailed discussions about this interpretation are given in Appendix C.

With the relationship between the $T_{\mathrm{ex}}$ and $T_{\mathrm{on,MPE}}$ in mind, we can examine the tunability of $T_{\mathrm{on,MPE}}$ by investigating its dependence on $t_{\mathrm{TmIG}}$ and choice of HM. Both W and Pt films exhibit increasing $T_{\mathrm{on,MPE}}$ with $t_{\mathrm{TmIG}}$. In the W/TmIG, $T_{\mathrm{on,MPE}}$ saturates at 7 nm (Fig. 1f), which is very long considering the interfacial nature of the exchange coupling. This $t_{\mathrm{TmIG}}$-dependent $T_{\mathrm{on,MPE}}$ is likely related to the TmIG saturation magnetization (see Appendix D). In the Pt/TmIG case, both 12 nm- and 15 nm-thick TmIG films yield $T_{on,MPE}$ above 380 K (see Appendix B). The higher $T_{on,MPE}$ in Pt for the same $t_{TmIG}$ is consistent with the fact that the Pt is closer to the Stoner instability and thus much easier to magnetize through proximity effect.

While the $T_{on,MPE}$ is always observed when the MPE presents, the AHE sign change does not always occur in the W/TmIG. We discuss this issue in Appendix E.

## B. MAGNETIZATION COMPENSATION TEMPERATURE

Having addressed the various AHE contributions, we note that in rare-earth transition metal alloys an AHE sign change has been observed across the $T_M$ since the spin polarization at the Fermi level is flipped across the $T_M$. Simultaneously, the $B_C$ diverges at $T_M$ since a zero-magnetization material is highly insensitive to an applied field. In contrast, the AHE response across $T_M$ in HM/FMI bilayer remains unclear since the Fermi level is in the bandgap of the FMI and no mobile carriers from the FMI contribute to the AHE. We explore this exchange coupling-induced AHE across the $T_M$ using W/TmIG. While in previous studies both bulk and thin-film TmIGs do not show a $T_M$ above 5 K [23, 24], some films in the present study exhibit a $T_M$ above 10 K. The presence and variability of $T_M$ is most likely due to cation off-stoichiometry, which is challenging to precisely control and may stabilize or boost the $T_M$ significantly even with small variation during growth. We experimentally identify this $T_M$ by investigating the $B_C$ of out-of-plane hysteresis loops (Fig. 2a). We observe a divergent $B_C$ around 75 K in a W/TmIG(6 nm) sample (Fig. 2b), the same temperature at which the AHE sign reverses, suggesting that the interfacial exchange coupling follows one sublattice rather than the net magnetization. We suspect that the exchange coupling effect follows the Fe sublattices since Fe *d*-orbitals are highly delocalized relative to Tm *f*-orbitals. We observe similar $T_M$-induced AHE sign changes and divergent $B_C$ in Pt/TmIG(6 nm), Pt/TbIG(6 nm) and W/TbIG (6 nm), where the $T_M$'s are 75 K, 290 K and 355 K, respectively (see Appendix B and F). Note that the Pt/TmIG(6 nm) and W/TmIG(6 nm) Hall bar devices are fabricated at different locations on the same TmIG thin film, so that the identical $T_M$ values strongly suggest that the $T_M$-induced AHE is insensitive to the choice of HM.

Highlighting the complex balance between all these effects, we note that two AHE sign changes occur in the same W/TmIG(6 nm) sample. As described above the AHE sign abruptly changes from negative to positive at 75 K, while the AHE sign gradually switches from positive to negative again near 45 K (Fig. 2b). At 75 K, we observe a divergent $B_C$ which identifies this transition as $T_M$, while the sign change at 45 K is accompanied by a relatively constant $B_C$. Further, removing the sign change associated with $T_M$ (Fig. 2b inset) by mirroring the AHE resistance below 75 K about the *x*-axis yield results in excellent agreement with those in Fig. 1e. Thus, we associate the sign change at 45 K with competition between MPE-AHE and SH-AHE.

## IV. XMCD

In order to confirm the validity of our analysis and demonstrate the usefulness of the AHE as a probe of both the HM and FMI, we examined the MPE and interfacial coupling using direct magnetization probes with elemental sensitivity and depth resolution. We employed XMCD, which uses circularly polarized photons and inherent spin-orbit coupling effects in electron energy level transitions to probe spin-dependent orbital occupancy and extract element-specific magnetic information from the W/TmIG. By tuning the incident X-ray energy to the resonant absorption edge of a given element and taking the absorption difference between left and right circularly polarized light, we may isolate the magnetization contribution of that element specifically. For XMCD measurements, we collected both total electron yield and luminescence yield data on $Nd_3Ga_5O_{12}$(111)/TmIG(10 nm)/W(5 nm)/Pt(2 nm) films. XAS spectra and XMCD were taken at beamline 4.0.2 of the advanced light source at a range of temperatures from 320 K to 8 K in applied fields of ±400 mT. Measurements were performed at the Fe $L^{3,2}$, Tm $M^5$, and W $N^3$ edges in the total electron yield and luminescence yield configurations at alternating applied fields and photon helicities.

X-ray absorption spectra (XAS) and XMCD taken at Fe $L_3$ edge and Tm $M_5$ edge are shown in Figs. 3a, b and Figs. 3c, d, respectively. The XMCD spectra reveal both Fe and Tm have a nonzero magnetization at all the investigated temperatures, but the magnetism of Tm exhibits a much stronger temperature dependence, nearly disappearing by 320 K (see Appendix G). This shows that Fe/W exchange coupling likely dominates over Tm/W, as expected. The XMCD spectra also show that the Fe and Tm have the opposite sign, indicating the two elements are anti-ferromagnetically coupled, consistent with previous studies [23] and as expected in most rare-earth iron garnets [25]. Although the extremely large $B_C$ near a $T_M$ necessitated measurements to be taken on a minor loop, we note that the Tm XMCD sign reverses through the suspected $T_M$ in one measured sample (see Appendix G).

XAS and XMCD measurements at W $N_3$ edge taken at 300 K and 80 K are shown in Figs. 3e, f and Figs. 3g, h, respectively. At 300 K, there is clearly no XMCD observed in the W, indicating an exceedingly weak MPE at higher temperatures. This indicates that the AHE above room temperature is due to the SHE. In contrast, a small but still distinguishable XMCD at the W $N_3$ edge appears at 80 K. We argue that the MPE-induced magnetic moment in the W is antiferromagnetically exchange-coupled to the Fe instead of the Tm (see inset in Fig. 3h) since Fe *d*-orbitals are relatively delocalized and Tm *f*-orbitals are more localized and previous studies have shown this antiferromagnetic exchange coupling in W/Fe systems [26, 27].

## V. PNR

To confirm the existence of a MPE in the W with antiparallel coupling, we utilize PNR to extract the magnetic and structural depth profile in a W/TmIG bilayer. For PNR measurements, we use $Nd_3Ga_5O_{12}$(111)/TmIG(10 nm)/W(5 nm)/$AlO_x$(3 nm). PNR measurements were performed after field cooling to 200 K and 80 K in an applied magnetic field of 700 mT using the PBR instrument at the NIST Center for Neutron Research. The measurement principle is discussed in Appendix H.

The best fits to the reflectivities and the resulting nuclear and magnetic scattering length density (SLD) profiles are shown in Fig. 4a and its inset. Here, the nuclear and magnetic SLDs are directly proportional to the nuclear scattering potential and the film magnetization respectively, so that fitting the data allows the structural and magnetic depth profiles to be deduced. The corresponding spin asymmetry and fit are shown in Fig. 4b. The PNR excludes the possibility of a MPE which couples ferromagnetically to the net Fe moment of the TmIG, instead favoring an antiparallel magnetization of 53(23) emu/cm$^3$ (1 emu/cm$^3$= 1 kA/m) at the interface at 200 K.

Similar results are obtained at 80 K. However, due to the huge perpendicular magnetic anisotropy effective field ($B_K \approx 2.8$ T, see Appendix I), the in-plane magnetization is very small. As shown in Fig. 5, we indeed observe that the measured magnetic moment is smaller and correspondingly, the measurement uncertainty is significantly larger than the case at 200 K. Nevertheless, qualitatively, the results are similar to those at 200 K, suggesting an antiparallel coupling between W and TmIG.

## VI. CONCLUSION AND DISCUSSION

In summary, both direct measurements of the magnetization, decomposing the magnetic signal as a function of element and depth within the film, reveal good agreement with the transport data and interpretations discussed above. Both PNR and XMCD favor the interpretation that the MPE favors antiparallel exchange coupling between the W and the Fe in the W/TmIG. Experimentally, we determine a positive MPE-AHE sign when the TmIG magnetization is pointing along the +*z* direction. To make a consistent comparison for different HMs, we define AHE sign in a magnetized HM when the HM magnetization is pointing along the +*z* direction. Since the measured MPE-AHE is positive and W and TmIG magnetizations are antiparallel, the magnetized W has a negative AHE sign. We now summarize the AHE sign associated with various magnetized HMs in Table I [6, 28]. With the information from the AHE, we can extract the exchange coupling configuration in arbitrary HM/magnetic insulator (MI) bilayers. For instance, Zhou *et al*. [6] and Amamou *et al*. [29] observed the AHE signs due to MPE are negative and positive for the Pd/YIG and Pt/$CoFe_2O_4$ ($CoFe_2O_4$ is a MI), respectively, so that we can predict parallel exchange coupling for both Pd/YIG and Pt/$CoFe_2O_4$ by using Table I. We also summarize results of the exchange coupling configurations in HM/magnet bilayers in Table II [6, 26, 27, 29-31], where all magnetic materials contain Fe elements. We can see that the exchange coupling configurations in HM/Fe bilayers are the same as in HM/MI bilayers, strongly suggesting that the exchange coupling is dominated by the HM-Fe exchange interaction. As discussed in [26, 27, 31], the exchange coupling configuration between two transition metals can typically be described using the Bethe-Slater curve, which describes the exchange coupling energy as a function of the ratio of the interatomic distance to the radius of the incompletely filled *d* shells. The ratio decreases when moving from the more to the less filled shells and leads to a sign change in exchange energy from positive (ferromagnetic) to negative (antiferromagnetic). The Pt and Pd have more-than-half-filled *d* shells, and thus a ferromagnetic exchange coupling, while W has less-than-half-filled *d* shells and thus an antiferromagnetic exchange coupling. The consistency of this picture is surprising considering the complexity of the oxide/metal interface. Note that future studies are encouraged to expand Table I and Table II.

*Note added*: We notice two very recent publications [32, 33] on the Pt/TbIG. Their results are consistent with ours and we analyze their data in our theoretical framework (see Appendix J).

## ACKNOWLEDGEMENTS

Qiming Shao would like to thank Guangyu Guo and Ran Cheng for helpful discussions. This work is supported as part of the Spins and Heat in Nanoscale Electronic Systems (SHINES), an Energy Frontier Research Center funded by the US Department of Energy (DOE), Office of Science, Basic Energy Sciences (BES) under award #S000686 and the National Science Foundation (DMR-1411085). We acknowledge the support from the Army Research Office Multidisciplinary University Research Initiative (MURI) program accomplished under Grant Number W911NF-16-1-0472 and W911NF-15-1-10561. This research used resources of the Advanced Light Source, which is a DOE Office of Science User Facility under contract no. DE-AC02-05CH11231. This work used the Extreme Science and Engineering Discovery Environment (XSEDE), which is supported by National Science Foundation grant number OCI-1053575. Specifically, it used the Bridges system, which is supported by NSF award number ACI-1445606, at the Pittsburgh Supercomputing Center (PSC). Certain commercial equipment is identified in this paper to foster understanding. Such identification does not imply recommendation or endorsement by NIST, nor does it imply that the materials or equipment identified are necessarily the best available for the purpose.

## APPENDIX A. AHE IN W/TmIG BILAYERS WITH DIFFERENT $t_{TmIG}$

In the main text, we present temperature dependent AHE resistance ($R_{AHE}$) in W(5 nm)/TmIG(15 nm) (Fig. 1e) and W(5 nm)/TmIG(6 nm) (Fig. 2b). In Fig. 6, we present the remaining $R_{AHE}$ data used to make Fig. 1f. Also, we present remaining $B_C$ data in Fig. 7.

We observe a nonmonotonic change of the $R_{AHE}$ slope below 100 K in W(5 nm)/TmIG(3.2 nm) (Fig. 6b), which is suggestive of a $T_M$. This is also indicated in the temperature dependence of $B_C$ in Fig. 7b, where the slope of the curve is nonmonotonic.

## APPENDIX B. AHE IN Pt/TmIG BILAYERS WITH DIFFERENT $t_{TmIG}$

In Fig. 8, we present the temperature dependent $R_{AHE}$ in Pt(5 nm)/TmIG($t_{TmIG}$) used to make Fig. 1f. Correspondingly, we present the temperature dependence of the $B_C$ in Fig. 9.

## APPENDIX C. INTERPRETING THE RELATION BETWEEN $T_{on,MPE}$ AND $T_{ex}$

In the main text, we interpret the local extrema of the AHE temperature dependence ($T_{ex}$), or the temperature at which AHE resistance slope sign reverses in Fig. 1e as an indicator of the MPE onset temperature ($T_{on,MPE}$). There are four primary reasons to draw this conclusion.

First, it has been predicted that in the absence of a MPE, the spin Hall effect (SHE)-induced AHE (SH-AHE) resistance is proportional to the magnetization $M$ [20]. We assume that, as suggested by room-temperature W XMCD, the MPE onset in our tungsten/thulium iron garnet (W/TmIG) samples is significantly below the Curie temperature of the MI ($T_{MI}$). This is unsurprising given that W is far from a Stoner instability and therefore difficult to magnetize. At the $T_{\mathrm{on,MPE}}$, the $M$ of the TmIG is nearly saturated since $M = M_0(1 - T/T_{\mathrm{MI}})^{\frac{1}{2}}$ and $T_{\mathrm{MI}} >> T_{\mathrm{on,MPE}}$. Thus, the SH-AHE is relatively insensitive to the temperature near the $T_{\mathrm{on,MPE}}$. In contrast, the MPE-induced AHE (MPE-AHE) should increase rapidly immediately below $T_{\mathrm{on,MPE}}$. Note that the exact temperature dependence of MPE-AHE may be very complex. In Fig. 10a, we summarize the temperature dependence of MPE-AHE resistance in graphene/YIG [14] and topological insulator (TI)/TmIG [16] from literature. We can see that they are very different from $\left(1 - T/T_{\mathrm{on,MPE}}\right)^{\frac{1}{2}}$ behavior. Empirically, the TI/TmIG data can be fit using a parabolic function. We obtain the theoretical curve in Fig. 10b, where we find that the parabolic temperature dependence assumption gives the most similar curve to the experimental data. Nevertheless, the $T_{\mathrm{ex}}$ is close to the $T_{\mathrm{on,MPE}}$.

Second, the presence of the MPE will suppress the SHE, as shown experimentally in ref. [22]. Therefore, the SH-AHE will be decreasing as the MPE-AHE develops. As the MPE becomes stronger with decreasing temperature, the SH-AHE will be suppressed further rather than increasing with the MI Magnetization, so that the SHE-AHE may even decrease. This makes the MPE-AHE more likely to dominate the SHE-AHE at low temperature, resulting in $T_{\mathrm{ex}}$.

Third, the $T_{\mathrm{ex}}$ increases as the TmIG thickness increases. This enhanced $T_{\mathrm{ex}}$ is consistent with the enhanced $T_{\mathrm{on,MPE}}$ as the TmIG saturation magnetization increases with the $t_{\mathrm{TmIG}}$ [17]. We discuss the MI thickness-dependent $T_{\mathrm{on,MPE}}$ in Appendix D.

Fourth, the $T_{\mathrm{ex}}$ is much higher in the Pt/TmIG than that in the W/TmIG at the same $t_{\mathrm{TmIG}}$. This is consistent with the fact that the Pt is much easier to magnetize as compared with the W since the Pt is closer to the Stoner instability.

## APPENDIX D. POSSIBLE MECHANISM FOR $t_{\mathrm{TmIG}}$-DEPENDENT $T_{\mathrm{on,MPE}}$

Here, we explore possible mechanism for achieving a MI thickness-dependent $T_{\mathrm{on,MPE}}$. The strength of the MPE in the HM/MI depends on both the magnetic susceptibility of the HM and surface (saturation) magnetization of the MI. (Typically, if the temperature is above the MI Curie temperature, there is no MPE since there is no magnetization.) We observe a much higher $T_{\mathrm{on,MPE}}$ for Pt than W at the same MI thickness, which is consistent with the fact that Pt has a much stronger susceptibility than W. We also observe that the $T_{\mathrm{on,MPE}}$ increases with the MI thickness with a characteristic length around 7 nm in W/TmIG, which is surprisingly large considering that the HM electrons cannot penetrate the MI over such long ranges. This could be explained by the thickness-dependent MI saturation magnetization, which saturates over a longer range. As shown in our TmIG thin films, the MI saturation magnetization and Curie temperature increases with the thickness and saturates around 10 nm (see Fig. 1c and 1d of ref. [17]) at room temperature. This contrasts with the saturation length around 1-2 nm for ferromagnetic metals (Co, CoFeB, etc.) at room temperature. Since the thicker MI film has a larger saturation magnetization at a given temperature, it provides a stronger exchange interaction (Fig. 11) and thus a higher $T_{on,MPE}$. The proof of our simple argument requires further theoretical and experimental investigations.

## APPENDIX E. DISCUSSION ON THE INTERMITTENT ABSENCE OF AHE SIGN CHANGE

In the low-temperature regime where the MPE is strongest, we expect an AHE sign change temperature ($T_1$) only if the MPE-AHE fully dominates over the SH-AHE. This sign change does not always happen in the W/TmIG as shown in Fig. 12. In the Pt/TmIG, we observe a $T_1$ in all the samples examined. However, there is no clear relation between $T_1$ and TmIG thickness in either the Pt/TmIG or W/TmIG. There are two possible explanations the lack of a $T_1$ in some W/TmIG. Firstly, it is possible that the $T_1$ occurs below 10 K, the lowest measured temperature, or that the coercive field is too large. Alternatively, we note that a $T_1$ requires that the MPE-AHE dominates over the SH-AHE. According to the theory [28], the MPE-AHE is very sensitive to the Fermi level position of the HM. For our 5 nm-thick W thin films, the resistivity varies from 140 to 170 μΩ·cm despite the use of same sputtering procedures and conditions. This variation in W may explain the absence of $T_1$ in some W/TmIG. Further investigations are required to clarify this point.

## APPENDIX F. AHE IN THE W/TbIG AND Pt/TbIG

To further validate our argument that across the $T_M$, the induced AHE in the HM layer changes sign, we probed the AHE in Pt/TmIG, Pt/TbIG and W/TbIG (Figs. 8 and 13) in addition to the W/TmIG. Note that the Pt is deposited on the same TmIG as the W in the W/TmIG series (3 nm, 6 nm, 9 nm, 12 nm and 15 nm). For each thickness, both Pt and W thin films occupy the half of surface of one 5mm × 5mm TmIG thin film before the Hall bar device fabrication. Consistently, only Pt on the 6 nm-thick TmIG shows a $T_M$, at which the AHE suddenly changes sign (Fig. 8b) and the $B_C$ diverges (Fig. 9b). Pt/TbIG and W/TbIG are prepared on different GGG substrates and both show a perpendicular magnetic anisotropy. The $T_M$ are 290 K and 355 K for Pt/TbIG and W/TbIG (Fig. 13), respectively, which are much higher than the bulk value (250 K). As expected, the AHE changes sign and the $B_C$ is divergent near the $T_M$ in these two bilayers as well.

## APPENDIX G. XMCD THROUGH THE $T_M$

Total electron yield and luminescence yield XMCD was taken for both the Fe L- and Tm M-edges through a suspected $T_M$. Unfortunately, the highest available field in the end station used was 400 mT, so that the magnetization could not be switched completely due to the divergence of the coercivity near the $T_M$. In this case, the Fe XMCD signal was too weak to clearly resolve. However, the Tm XMCD remained measurable and its temperature dependence is plotted in Fig. 14. Even measurements along a minor magnetization hysteresis loop provide significant insight, and in this case the XMCD on the Tm edge is reversed below the suspected $T_M$, confirming our interpretation of $T_M$ in some of our TmIG thin films.

## APPENDIX H. PNR PRINCIPLE

Measurements were performed in the specular reflection geometry, with the direction of wave vector transfer perpendicular to the film surface. The neutron propagation direction was perpendicular to both the sample surface and the applied field direction. In any case, the perpendicular anisotropy of TmIG ensures that moments which do not align fully along the in-plane field will instead cant along the growth axis and consequently will not produce spin-flip scattering. We therefore consider only the non-spin-flip scattering cross sections and in all cases the incident and scattered neutrons were polarized either spin-up or spin-down with respect to the applied magnetic field. Scattering length density (SLD) is a measure of the potential experienced by the neutron as a function of depth within the sample. Specifically, if we define the potential energy of a neutron traveling in a given medium as $V$, then the nuclear SLD (associated with scattering from nuclei) is linearly related to the potential by

$$V = \frac{2\pi\hbar^2}{m} SLD_{Nuclear}$$

While the magnetic SLD is simply an adjustment which depends on the magnetization of the media and the direction of the neutron spin. Specifically,

$$SLD_{Magnetic} = \mp \frac{m}{2\pi\hbar^2} \mu B$$

Where the sign depends on the neutron spin direction, B is the magnetic field, and μ is the neutron magnetic moment. The nuclear and magnetic SLDs are directly proportional to the nuclear scattering potential and the film magnetization respectively, so that fitting the data allows the structural and magnetic depth profiles to be deduced. The reflected intensity was measured as a function of the momentum transfer vector *Q* and modeled using the NIST Refl1D software package [34].

## APPENDIX I. TEMPERATURE DEPENDENT $B_K$

To quantify the in-plane magnetization component when we subject the sample to a 700 mT in-plane external field in polarized neutron reflectometry (PNR) experiments, we determine the $B_K$ at different temperatures for a reference sample W(5 nm)/TmIG(10 nm) using hard-axis (in-plane) Hall hysteresis loops (Fig. 15). The determined $B_K$'s are 470 mT and 2.8 T at 200 K and 80 K, respectively.

## APPENDIX J. DISCUSSION ON THE RECENT TWO PUBLICATIONS ON THE Pt/TbIG

In two recent publications [32, 33], the AHE temperature dependence in the Pt/TbIG were reported. We plot these data in Fig. 16. Fig. 16a reveals a $T_M$ around 230 K and a $T_{on,MPE}$ around 140 K. Fig. 16b reveals a $T_M$ around 355 K and a $T_{on,MPE}$ higher than 350 K. We show that their data can be interpreted using our temperature-dependent AHE model, although the details and parameters may vary somewhat.

**Figures and captions**

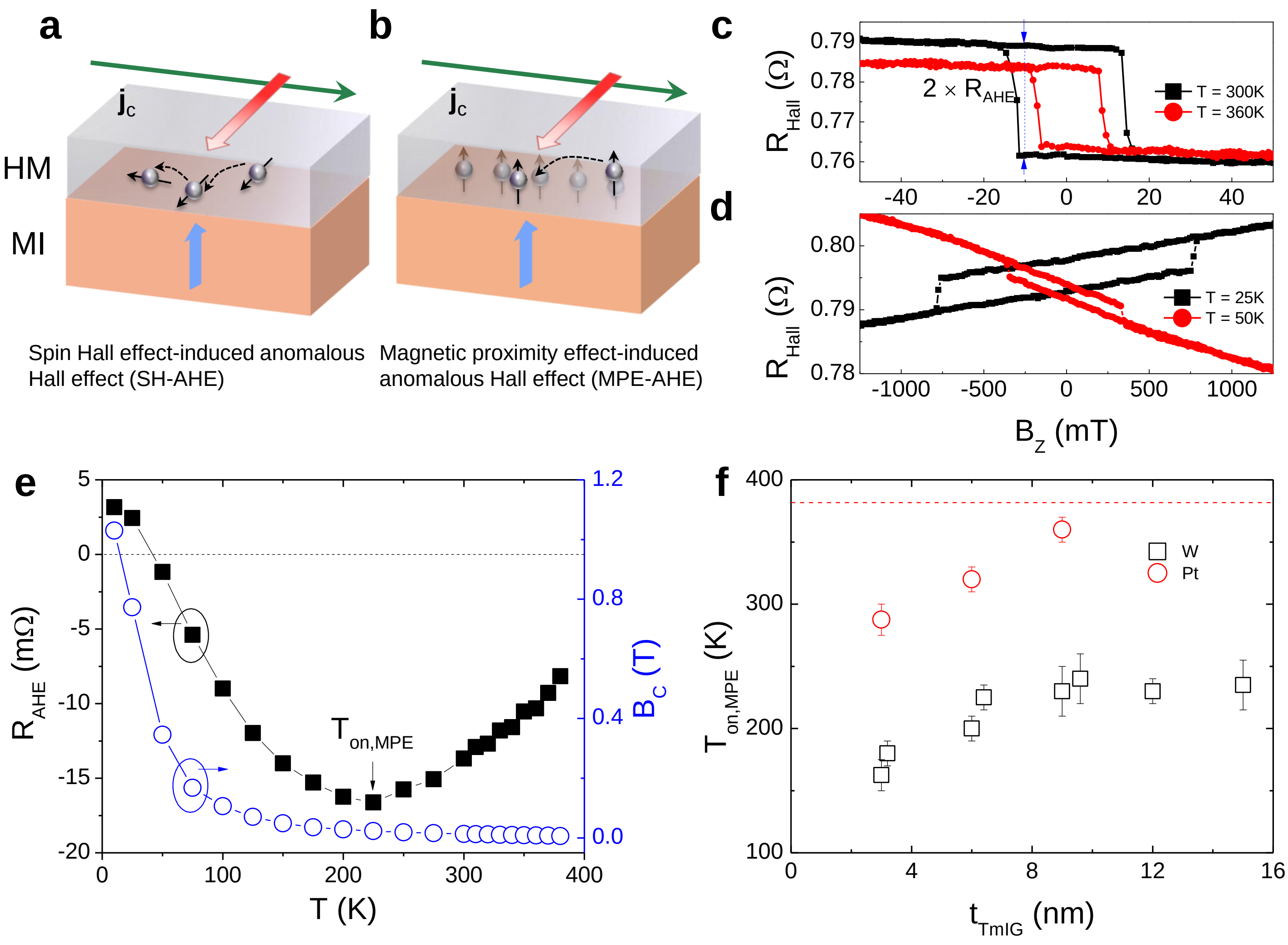

**Figure 1**. Temperature dependent AHE in HM/TmIG. (a-b) Schematics of SH-AHE and MPE-AHE, respectively, in HM/magnetic insulator heterostructures. For the SH-AHE, the reflected spin angular momenta are rotated by 90 degrees compared with the incident spin angular momenta due to spin-dependent scattering at the interface. This rotated spin angular momenta create a transverse charge current due to inverse SHE, resulting in an AHE. For the MPE-AHE, the AHE is from the interfacial magnetized HM layer due to the MPE. (c-d) Hall resistance as a function of out-of-plane magnetic field for $T$ = 300 K and 360 K (c) and $T$ = 25 K and 50 K (d) for a W(5 nm)/TmIG(15 nm) bilayer. (e) AHE resistance and coercive field of out-of-plane hysteresis loops as a function of temperature for a W(5 nm)/TmIG(15 nm) bilayer. MPE onset temperature is indicated by the arrow $T_{on,MPE}$. (f) Onset temperature as a function of TmIG thickness in both the W/TmIG and Pt/TmIG. The error bars reflect standard deviations from multiple measurements.

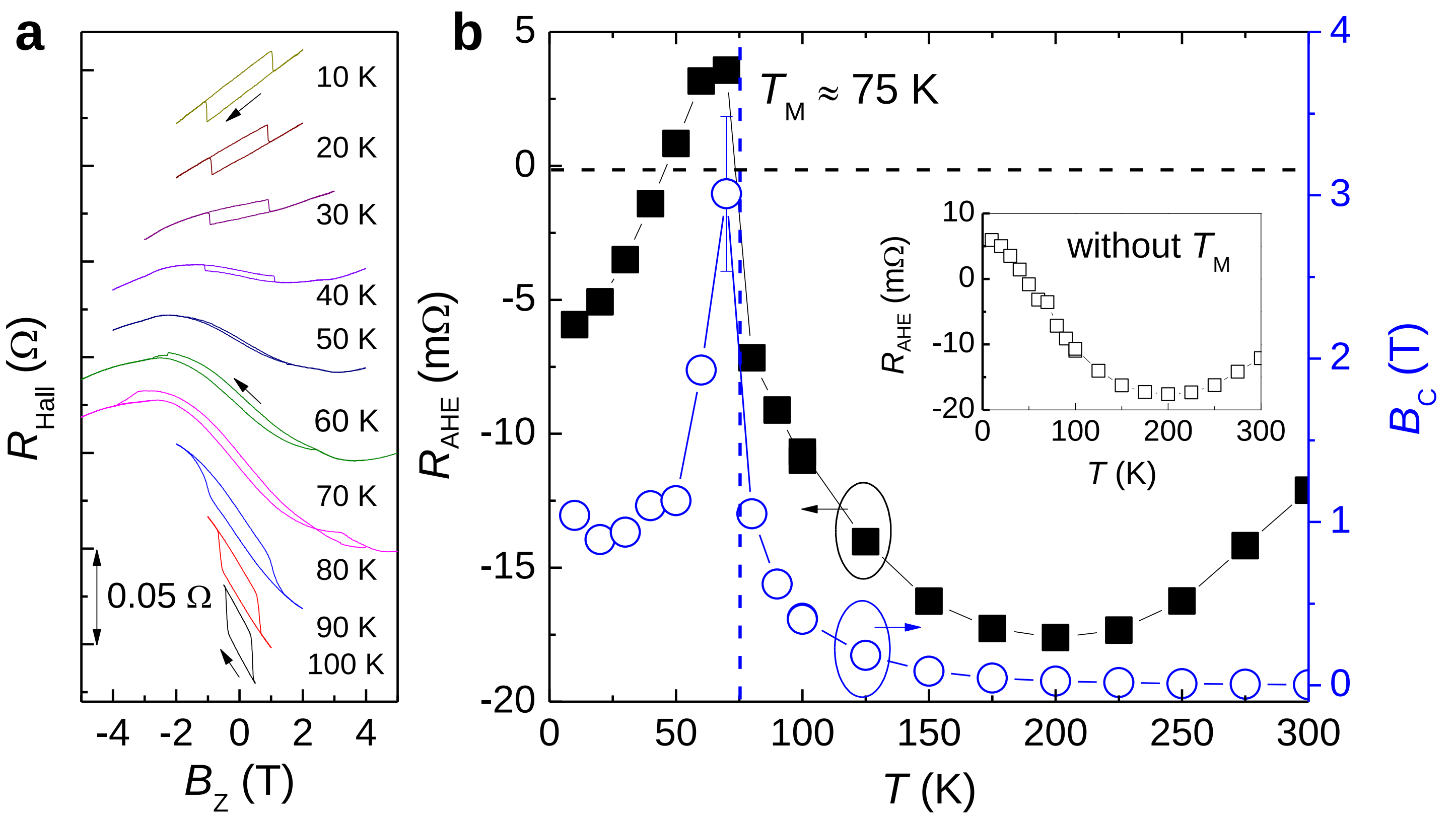

**Figure 2**. Emergence of the AHE sign change at the magnetization compensation temperature ($T_M$) in a W(5 nm)/TmIG(6 nm) bilayer. (a) Hall resistance vs. out-of-plane magnetic field for different temperatures. The arrow indicates the field sweeping direction. (b) AHE resistance and coercive field of out-of-plane hysteresis loops as a function of temperature. The vertical blue dashed line indicates the $T_M$. Inset is the inferred data for the case without a $T_M$.

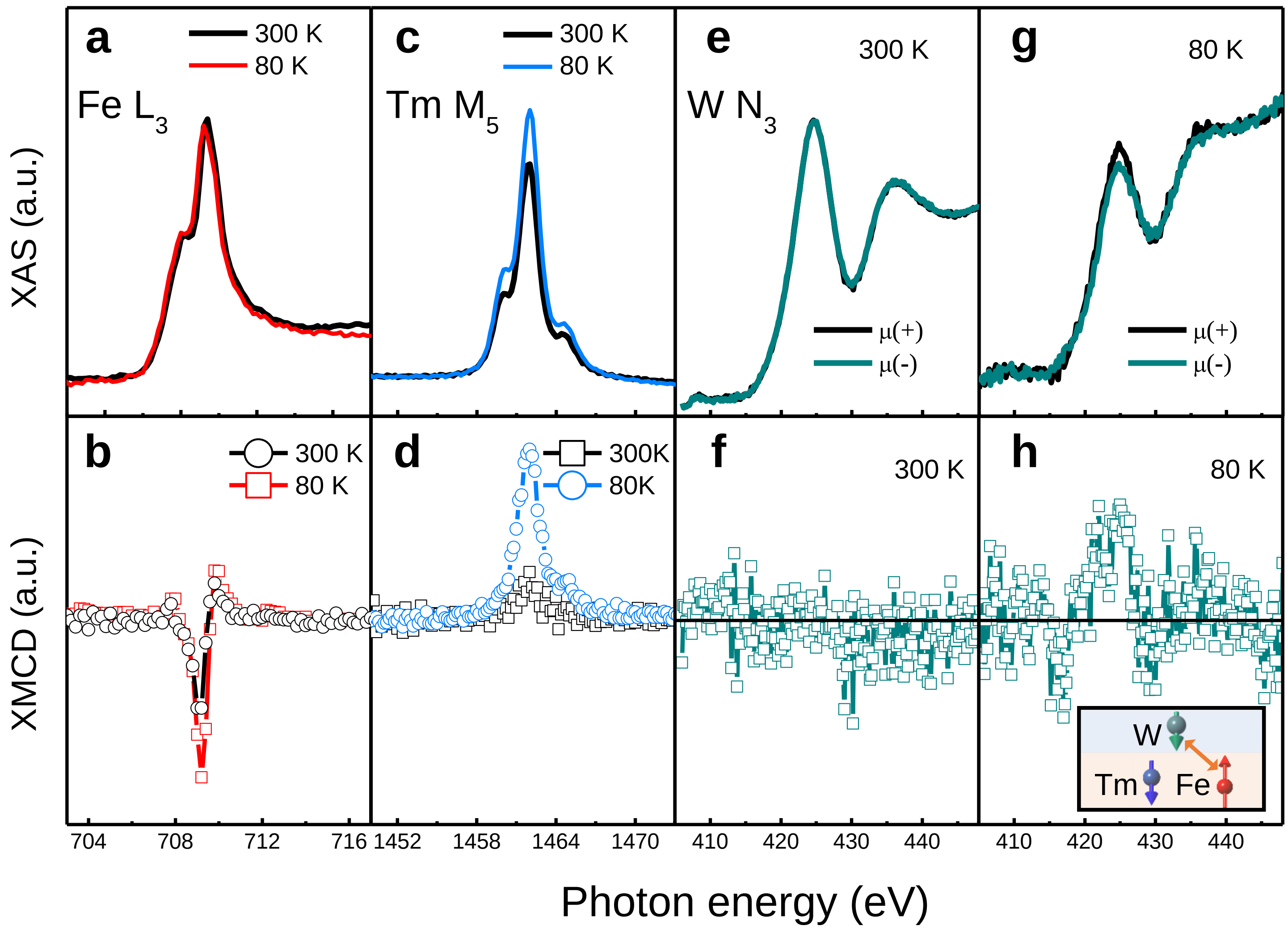

**Figure 3**. Capturing the exchange interactions in the W(5 nm)/TmIG(10 nm) by X-ray techniques. (a) XAS and (b) XMCD spectra taken at Fe $L_3$ edge at 80 K and 300 K. (c) XAS and (d) XMCD spectra taken on Tm $M_5$ edge at 80 K and 300 K. XAS taken on W $N_3$ edge at 300 K (e) and 80 K (g) with two opposite x-ray helicities, μ(+) and μ(-). XMCD at W $N_3$ edge taken at 300 K (f) and 80 K (h). Inset in (h) illustrates relative spin alignments of the Fe, Tm, and induced W moment at 80 K based on the sign of XMCD.

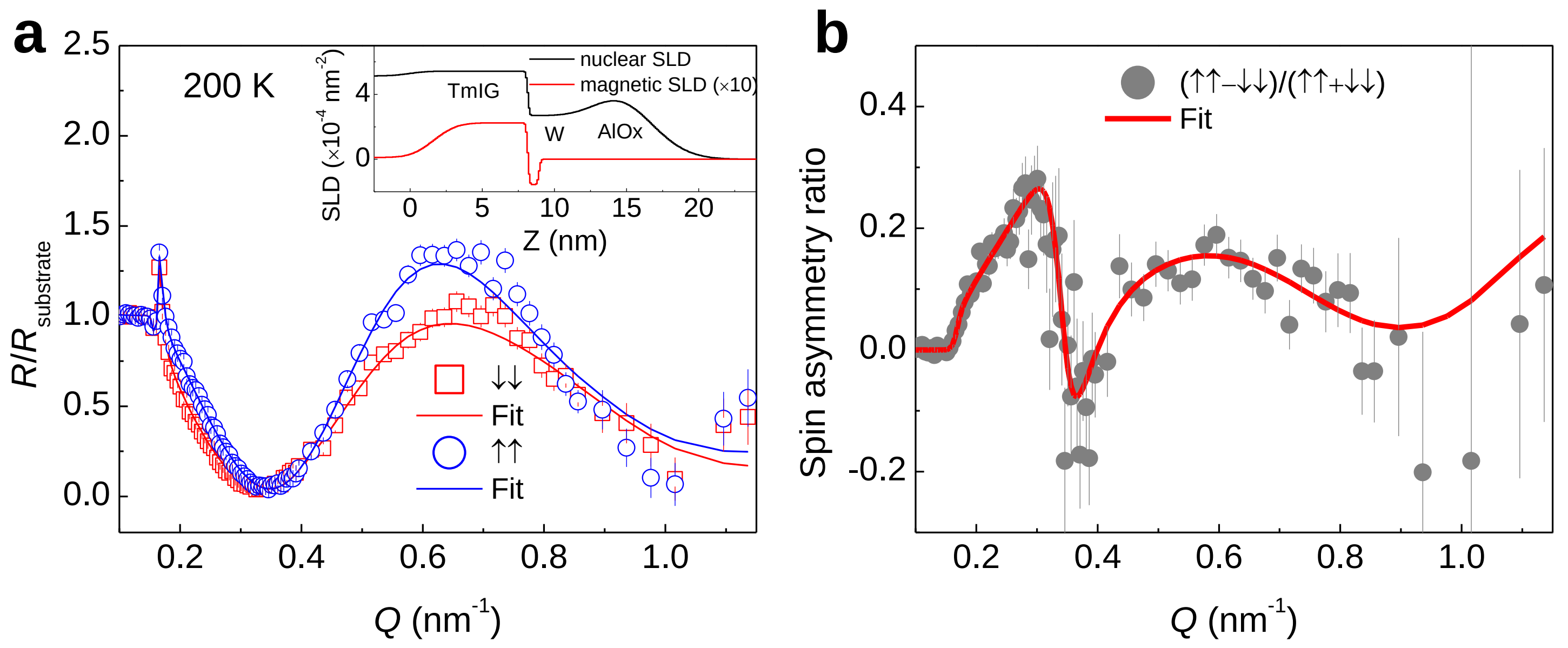

**Figure 4**. Capturing the spin textures in the W(5 nm)/TmIG(10 nm) by neutron techniques at 200 K. (a) Polarized neutron reflectivities (with a 700 mT in-plane field) for the spin-polarized $R^{\uparrow\uparrow}$ and $R^{\downarrow\downarrow}$ channels. Inset shows the corresponding models with structural and magnetic scattering length densities (SLDs) used to obtain the best fits. (b) The spin asymmetry ratio $(R^{\uparrow\uparrow} - R^{\downarrow\downarrow})/\ (R^{\uparrow\uparrow} + R^{\downarrow\downarrow})$ between the $R^{\uparrow\uparrow}$ and $R^{\downarrow\downarrow}$ channels. The error bars are ±1 s.d.

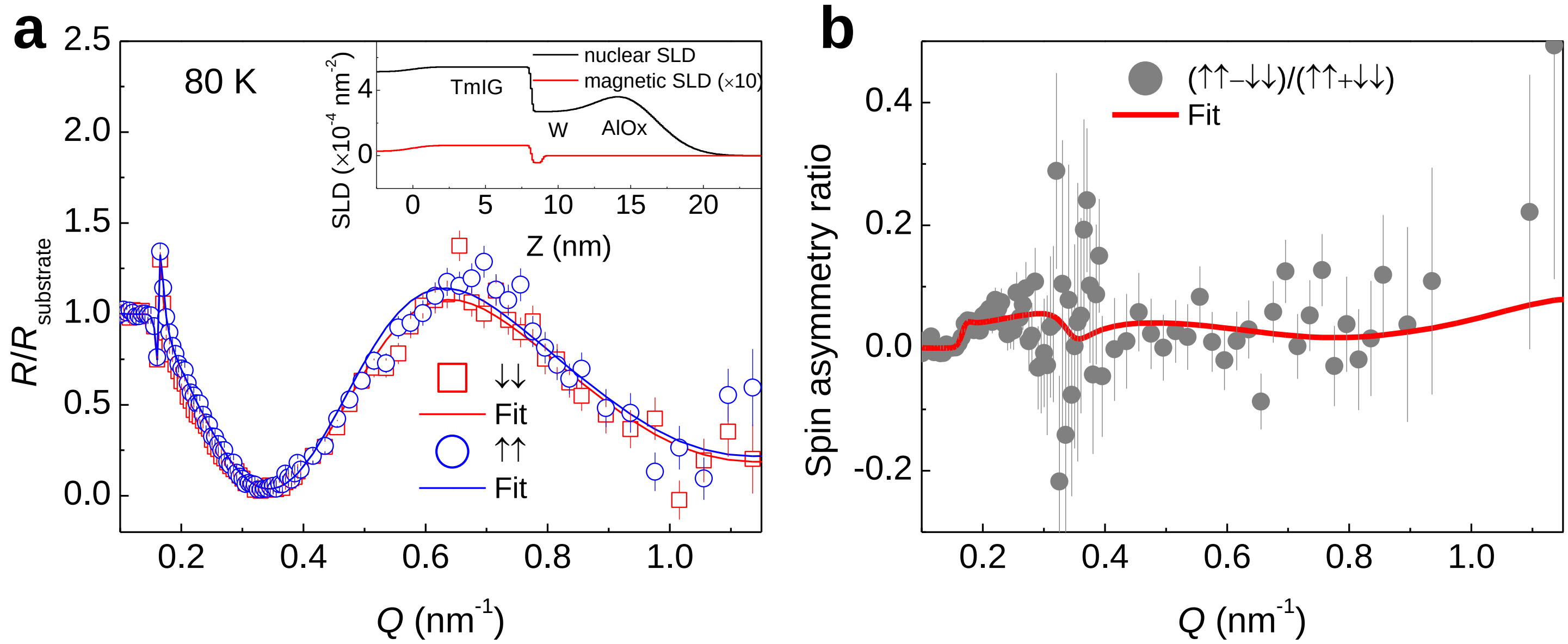

**Figure 5**. Capturing the spin textures in the W(5 nm)/TmIG(10 nm) by neutron techniques at 80 K. (a) Polarized neutron reflectivities (with a 700 mT in-plane field) for the spin-polarized $R^{\uparrow\uparrow}$ and $R^{\downarrow\downarrow}$ channels. Inset shows the corresponding models with structural and magnetic scattering length densities used to obtain the best fits. (b) The spin asymmetry ratio $(R^{\uparrow\uparrow} - R^{\downarrow\downarrow})/(R^{\uparrow\uparrow} + R^{\downarrow\downarrow})$ between the $R^{\uparrow\uparrow}$ and $R^{\downarrow\downarrow}$ channels. The error bars are ±1 s.d.

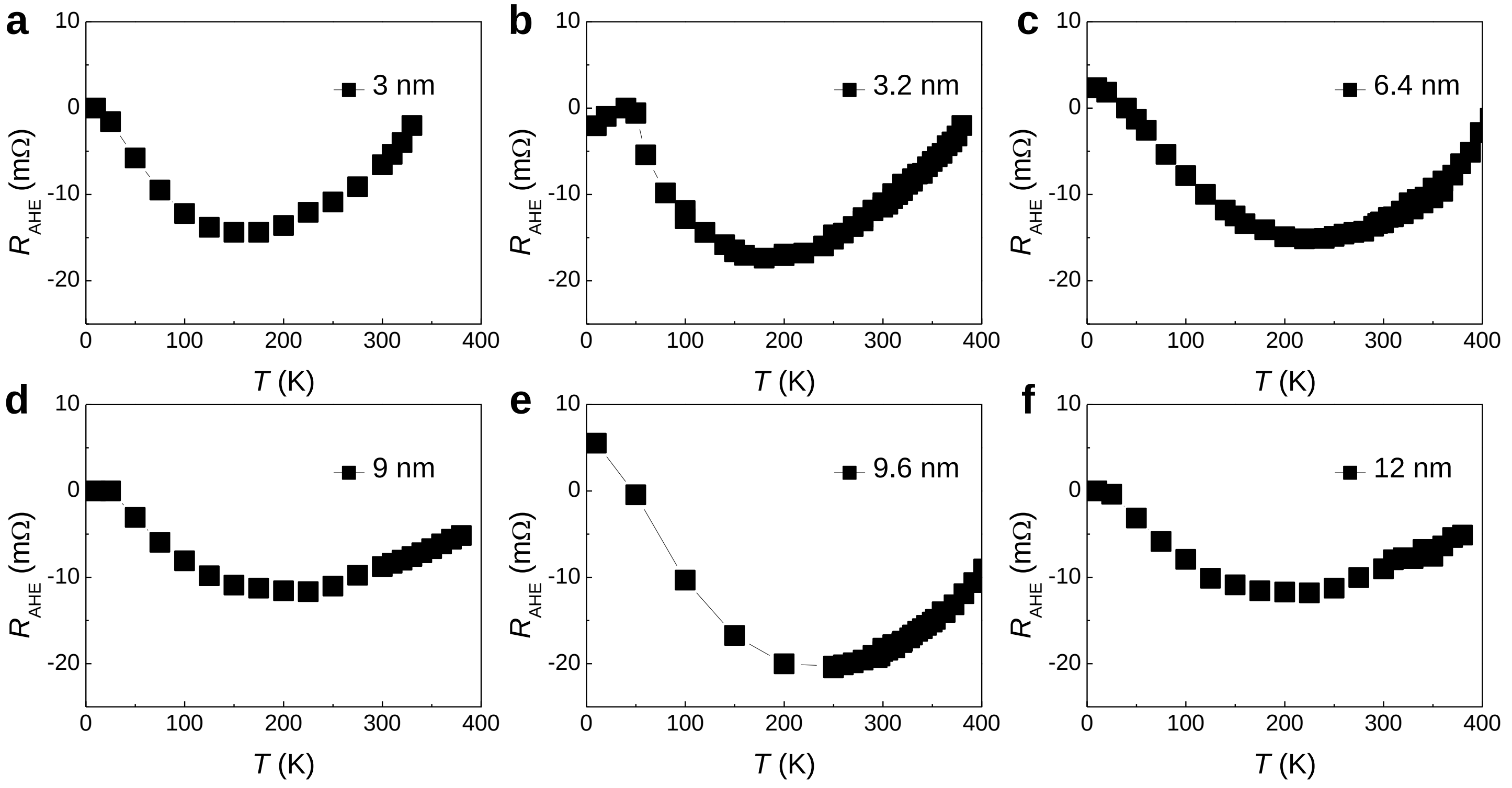

**Figure 6**. Temperature dependence of $R_{\mathrm{AHE}}$ in W/TmIG bilayers with different TmIG thickness.

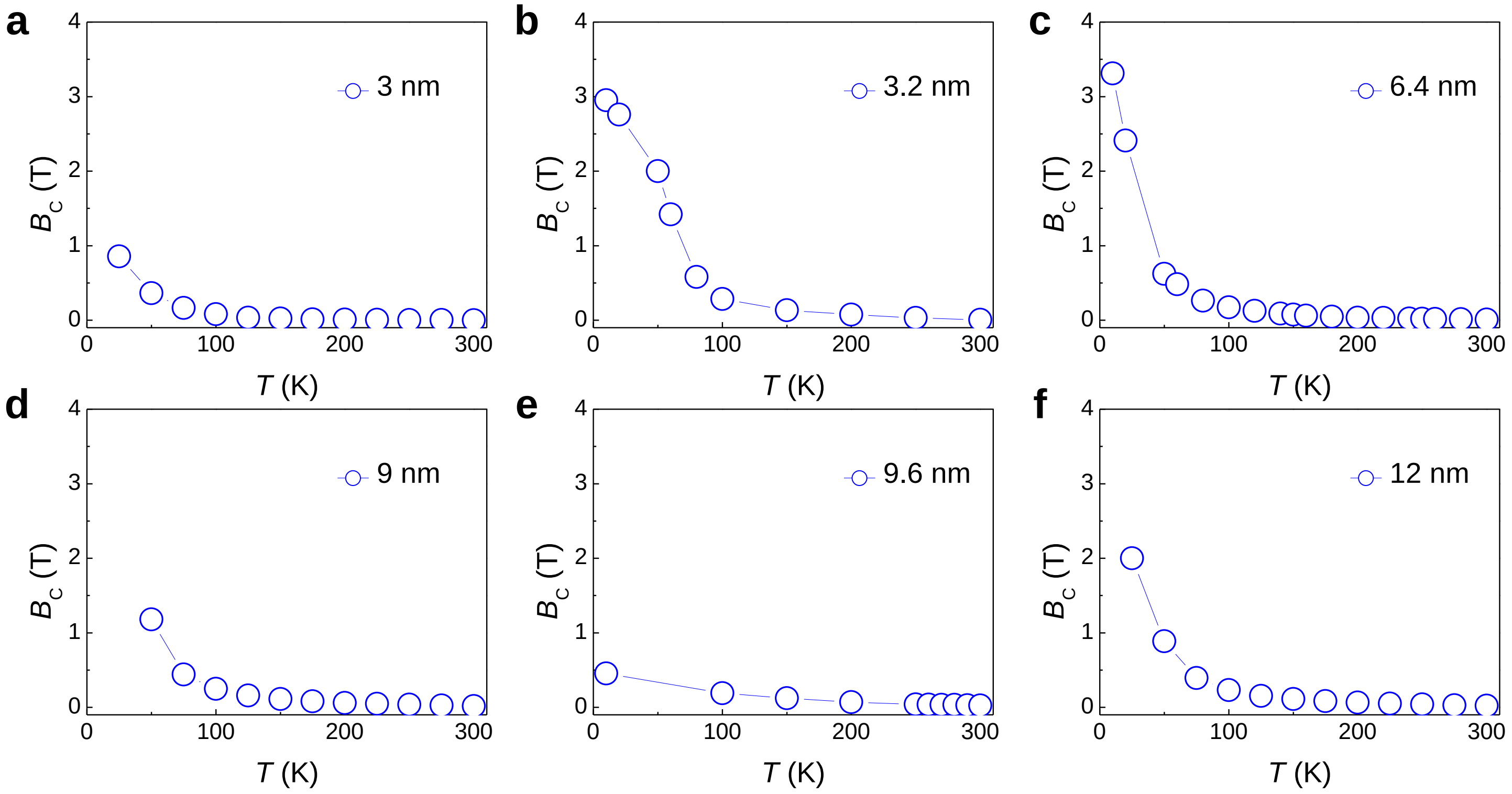

**Figure 7**. Temperature dependence of $B_C$ in W/TmIG bilayers with different TmIG thickness.

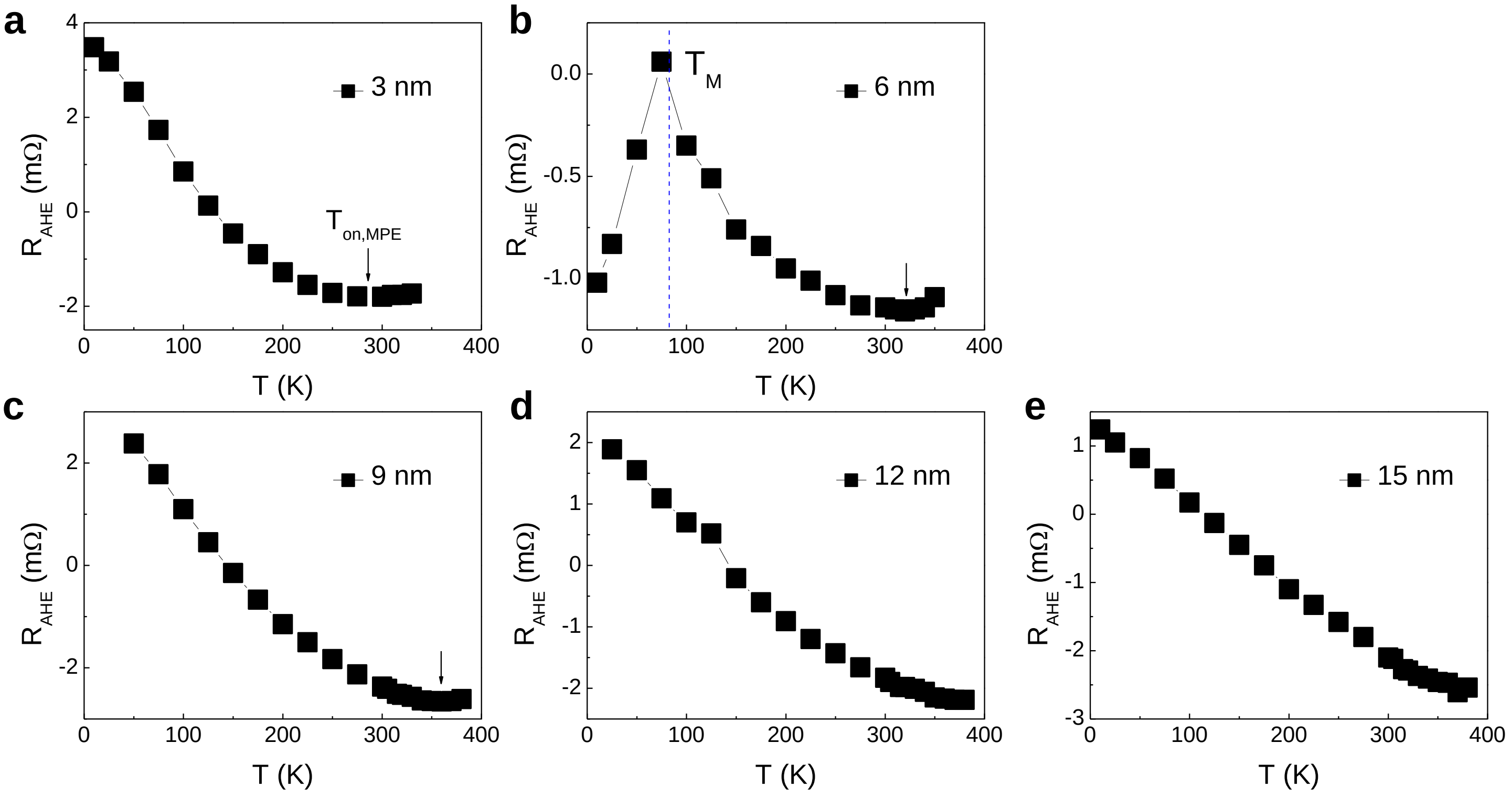

**Figure 8**. Temperature dependence of $R_{AHE}$ in Pt/TmIG bilayers with different TmIG thickness.

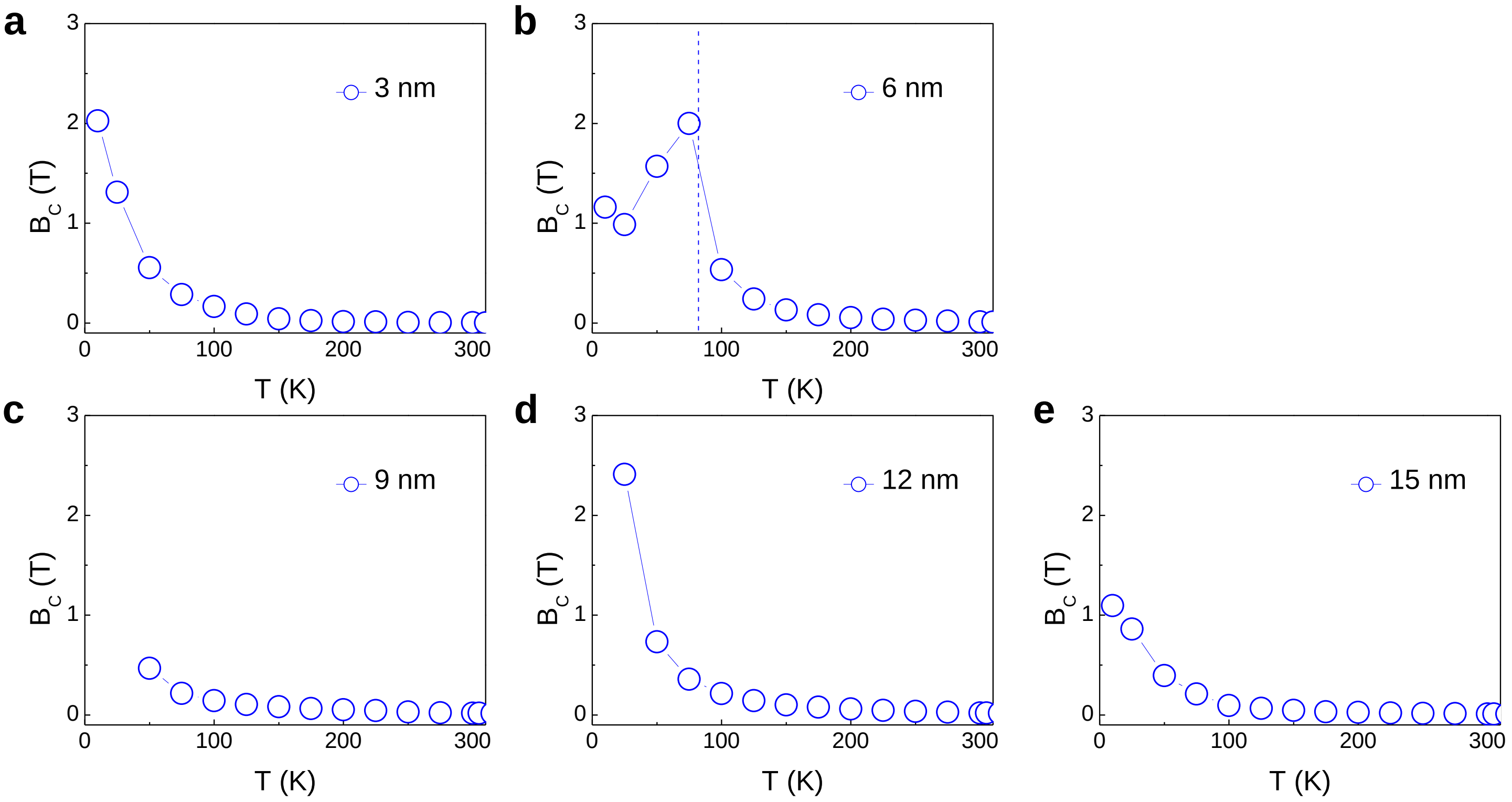

**Figure 9.** Temperature dependence of $B_C$ in Pt/TmIG bilayers with different TmIG thickness.

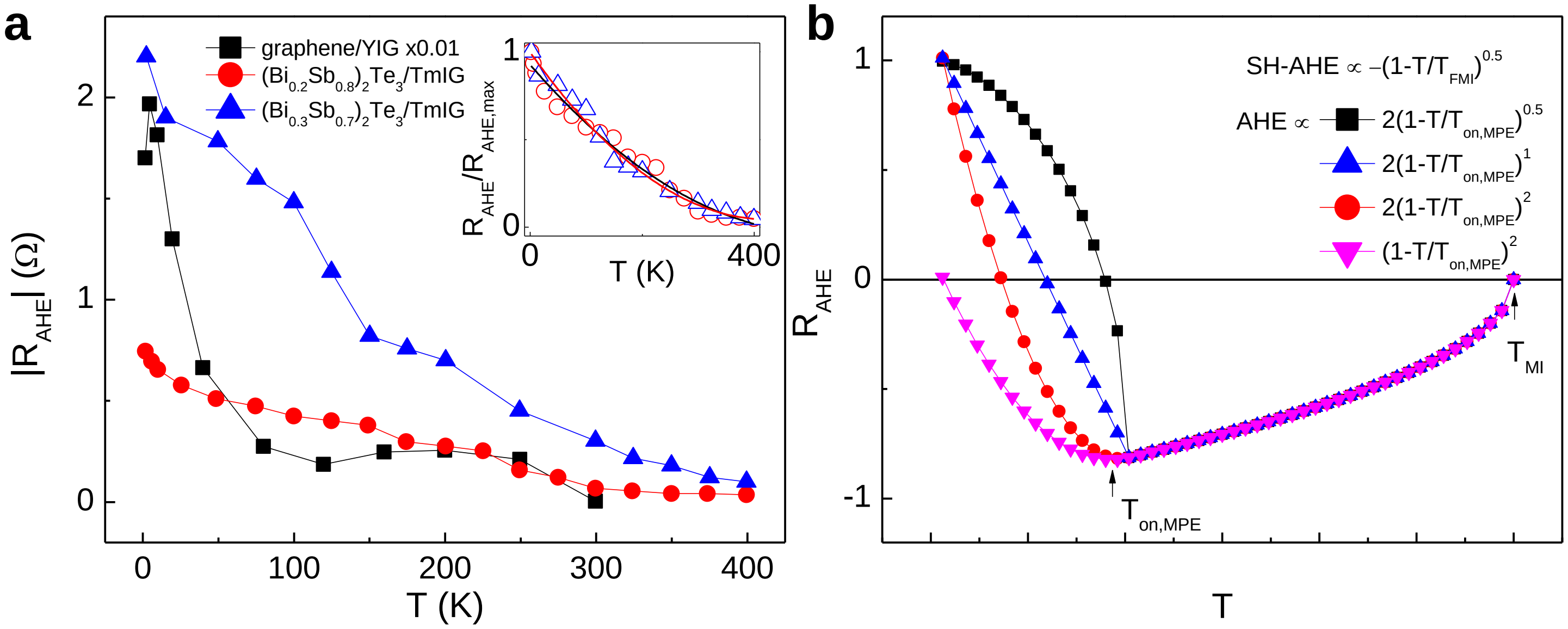

**Figure 10**. (a) Temperature dependence of AHE resistance in graphene/YIG [14] and TI/TmIG [16] systems. The inset shows parabolic fitting to the normalized AHE resistance data of TI/TmIG. (b) Schematic of AHE resistance due to competition between MPE-AHE and SH-AHE. Temperature dependences of MPE-AHE with different scaling exponents and coefficients are shown.

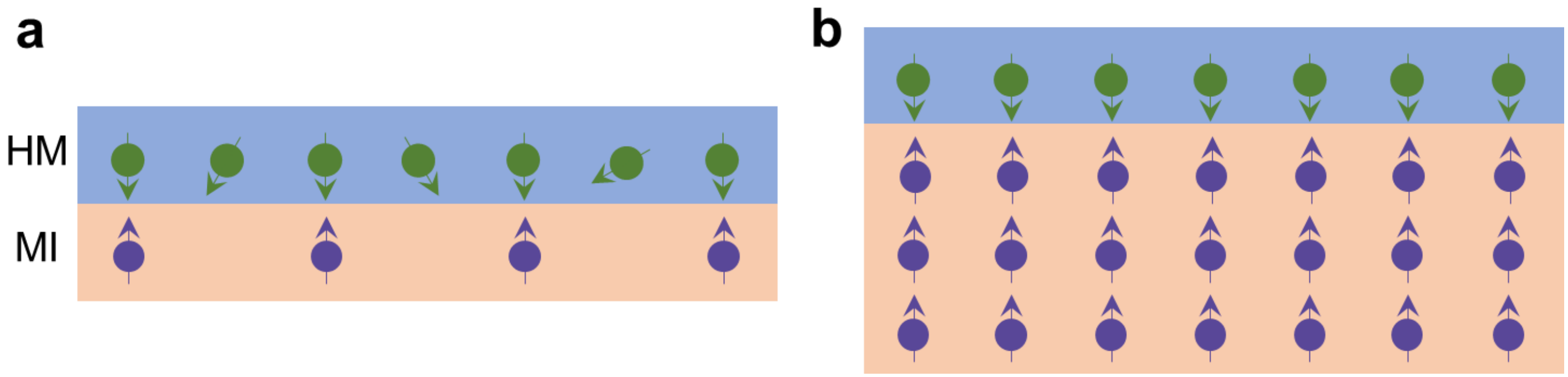

**Figure 11**. Schematic of exchange coupling at finite temperature in the HM/MI bilayer. Purple arrows represent the atomic magnetic moments in the MI, whose density represents the saturation magnetization. The surface HM atoms (green arrows) interact with the surface magnetization of MI. When the MI is much thinner like in case (a) than the bulk case (b), the $T_{MI}$ is strongly suppressed and thus at a finite temperature (around the half of the MI Curie temperature), the saturation magnetization is much smaller in (a) than (b). Smaller saturation magnetization leads to weaker exchange interaction and thus lower $T_{on,MPE}$.

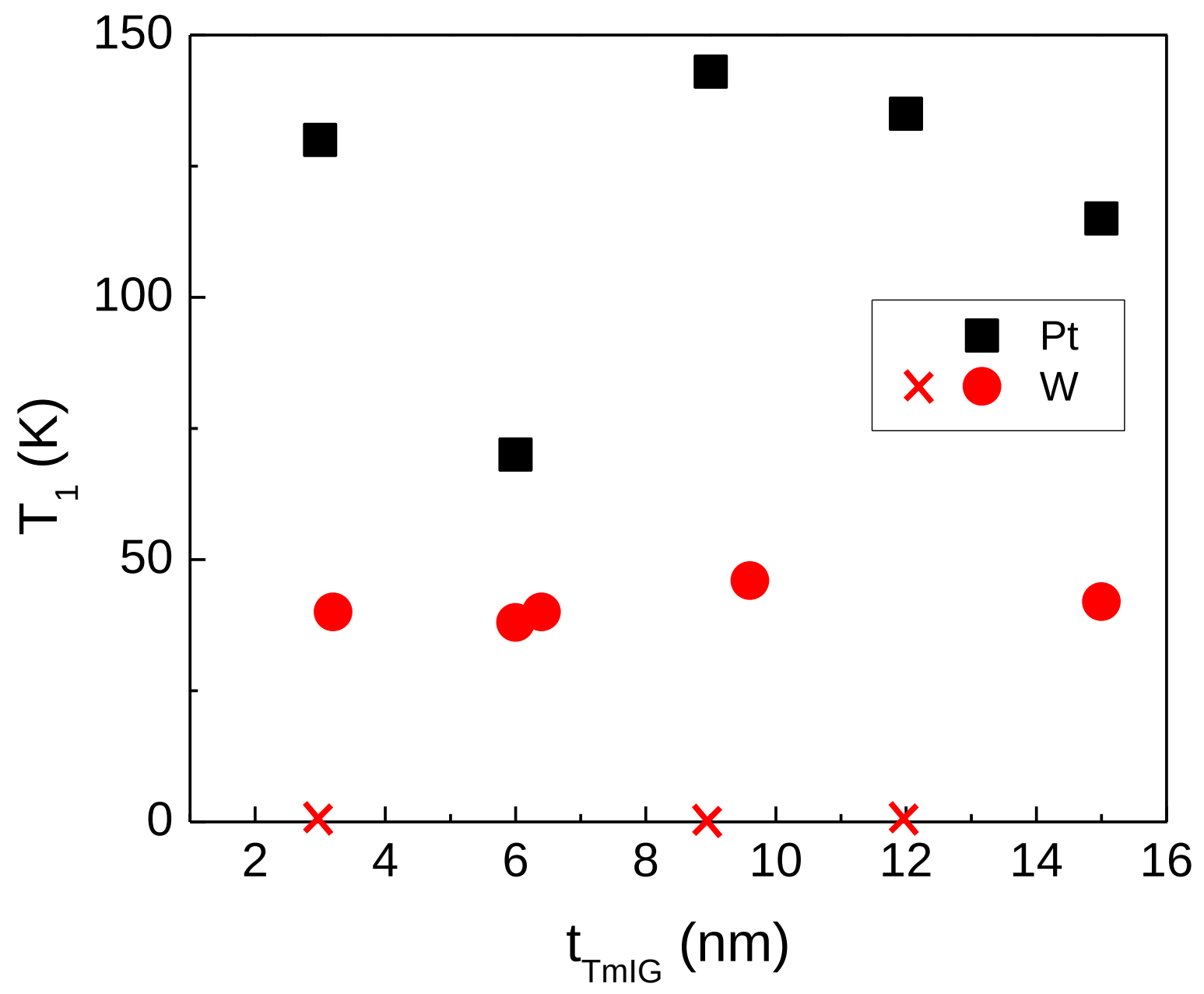

**Figure 12.** The low temperature AHE sign change temperature ($T_1$) due to the MPE in the Pt/TmIG and W/TmIG with different TmIG thicknesses. The label × on the *x*-axis indicates that the $T_1$ is not clearly observed.

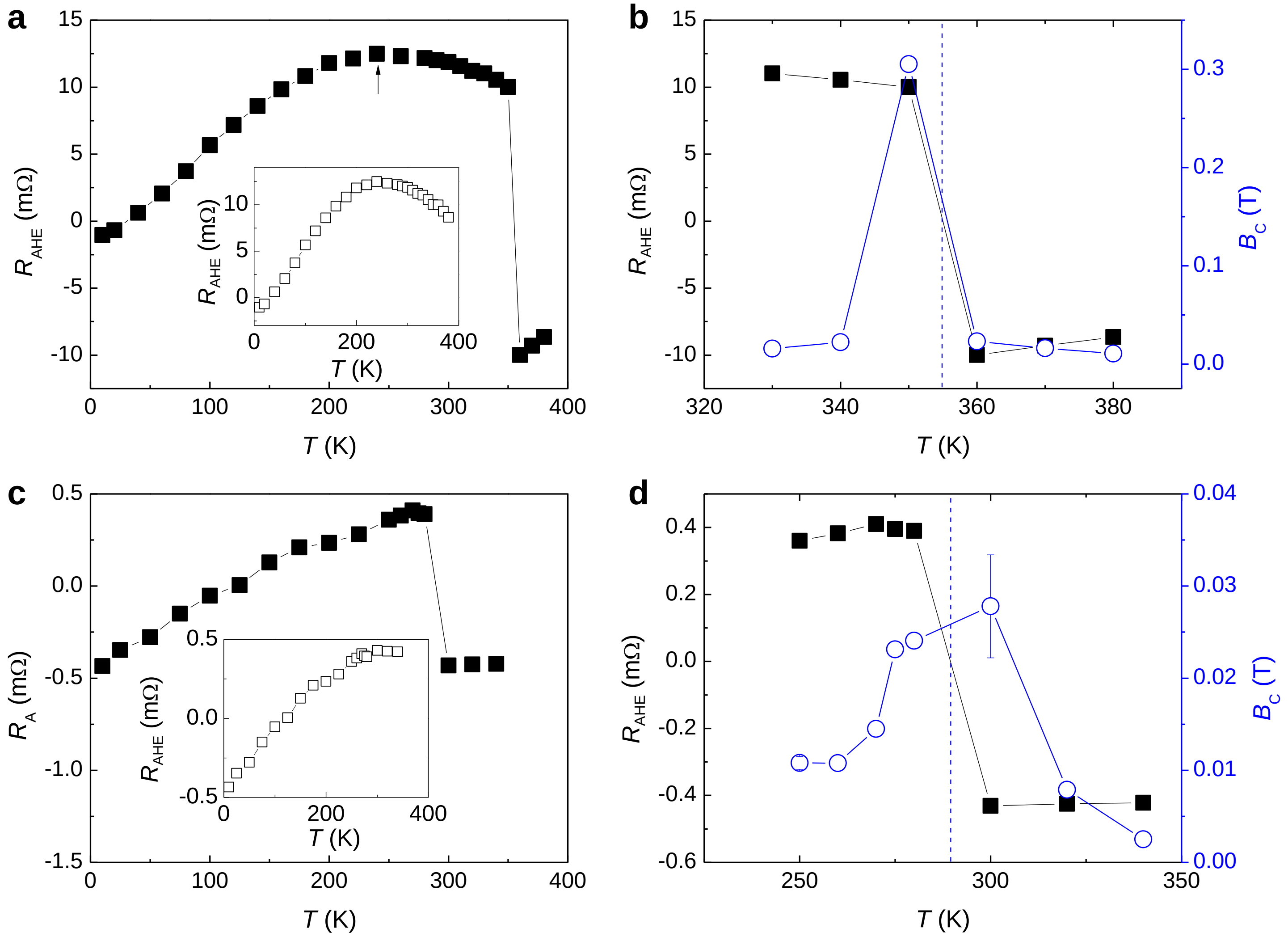

**Figure 13.** (a) AHE resistance as a function of temperature in the W(5 nm)/TbIG(6 nm). Inset is the inferred data for the case without a $T_M$. (b) AHE resistance and coercive field of out-of-plane hysteresis loops near the $T_M$ in the W(5 nm)/TbIG(6 nm). The vertical blue dashed line indicates the $T_M$. (c) AHE resistance as a function of temperature in the Pt(5 nm)/TbIG(6 nm). Inset is the inferred data for the case without a $T_M$. (d) AHE resistance and coercive field of out-of-plane hysteresis loops near the $T_M$ in the Pt(5 nm)/TbIG(6 nm). The vertical blue dashed line indicates the $T_M$.

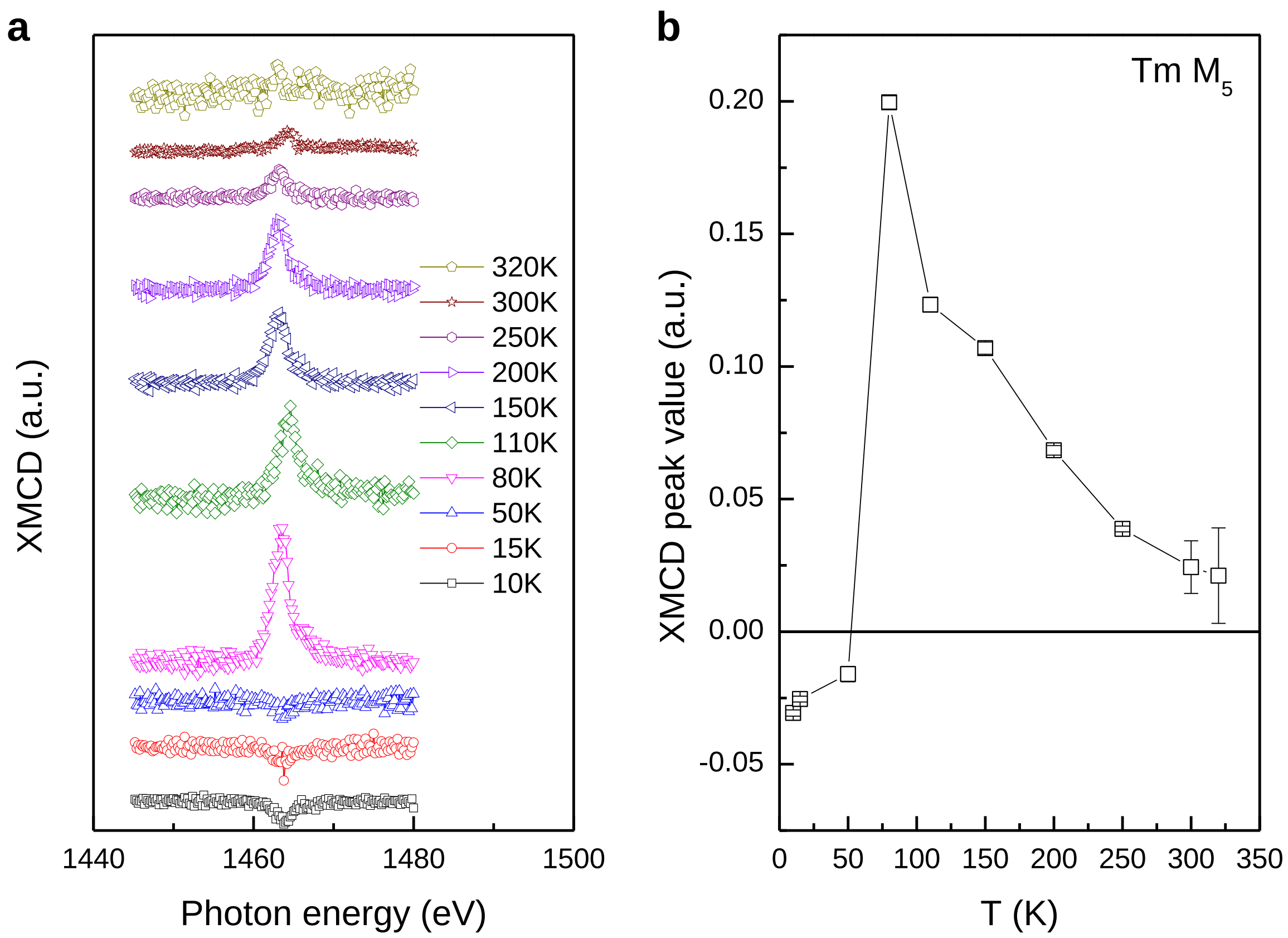

**Figure 14.** (a) XMCD signals at different temperature. (b) Tm $M_5$ XMCD peak value as a function of temperature.

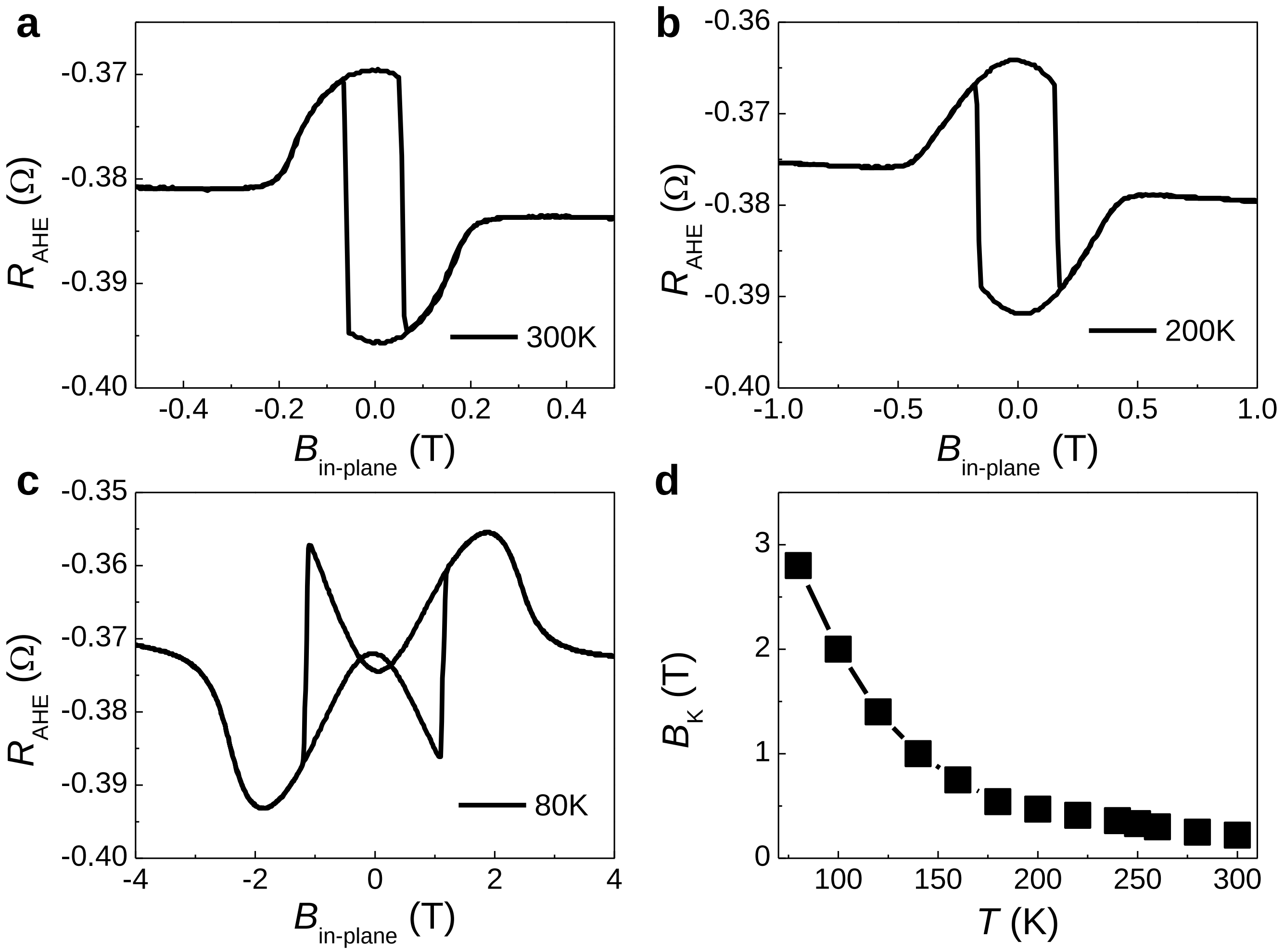

**Figure 15**. Temperature dependence of in-plane Hall hysteresis loops at 300 K (a), 200 K (b) and 80 K (c). (d) Temperature dependence of $B_K$.

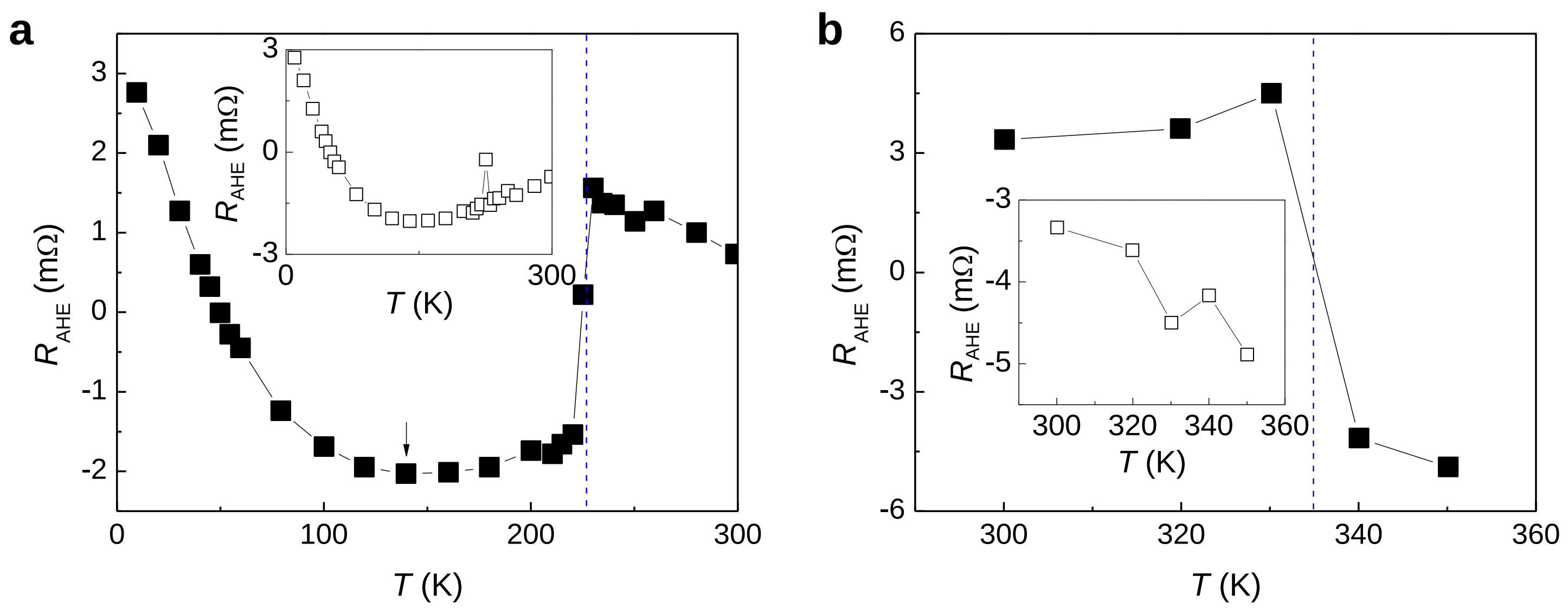

**Figure 16.** Temperature dependence of AHE in the Pt(5nm)/TbIG(30nm) [32] and the Pt(4nm)/TbIG(10nm) [33].

**Table I. Sign of AHE in various magnetized heavy metals**

| Heavy metal element | Pt | Pd | W |
|---|---|---|---|
| Sign of AHE | Positive (this work and [6, 28]) | Negative[6, 28] | Negative (this work) |

**Table II. Exchange coupling configuration in various heavy metal/magnet bilayers**

| Type of magnet | Magnetic metal | | | Magnetic insulator | | |
|---|---|---|---|---|---|---|
| Bilayer structure | Pt/ Fe [26, 27] | Pd/ Fe [27, 31] | W/ Fe [26, 27] | Pt/ $Y_3Fe_5O_{12}$ [30], $CoFe_2O_4$ [29], $Tm_3Fe_5O_{12}$*, $Tb_3Fe_5O_{12}$*) | Pd/ $Y_3Fe_5O_{12}$ [6] | W/ ($Tm_3Fe_5O_{12}$*, $Tb_3Fe_5O_{12}$*) |
| Exchange coupling configuration | FM | FM | AFM | (FM, FM**, FM**, FM**) | FM* | (AFM, AFM**) |

* This work

** Predicted using the experimental AHE sign and Table I